\documentclass[a4paper,UKenglish]{lipics-v2016}
\newif\ifFull
\Fullfalse
\usepackage{booktabs} 

\newif\ifFull
\Fullfalse
\usepackage{t1enc}
\usepackage[utf8]{inputenc}

\usepackage{amsfonts}

\newcommand{\set}[1]{\left\{ #1\right\}}
\newcommand{\gilt}{:}
\newcommand{\sodass}{\,:\,}
\newcommand{\setGilt}[2]{\left\{ #1\sodass #2\right\}}

\newcommand{\realrange}[2]{\left[#1, #2\right]}

\newcommand{\unitrange}[2]{\realrange{0}{1}}

\newcommand{\llabel}[1]{\label{\labelprefix:#1}}
\newcommand{\labelprefix}{} 

\newcommand{\discussionsize}{\small}

\newenvironment{code}{\noindent
\begin{tabbing}%
\hspace{2em}\=\hspace{2em}\=\hspace{2em}\=\hspace{2em}\=\hspace{2em}\=%
\hspace{2em}\=\hspace{2em}\=\hspace{2em}\=\hspace{2em}\=\hspace{2em}\=%
\kill}{\end{tabbing}}

\newcommand{\labelcommand}{}
\newcommand{\captiontext}{}
\newsavebox{\codeparam}
\newcounter{lineNumber}
\newenvironment{disscodepos}[3]{%
\renewcommand{\labelcommand}{#2}%
\renewcommand{\captiontext}{#3}%
\sbox{\codeparam}{\parbox{\textwidth}{#3}}%
\begin{figure}[#1]\begin{center}\begin{code}\setcounter{lineNumber}{1}}{%
\end{code}\end{center}\caption{\llabel{\labelcommand}\captiontext}\end{figure}}

{\end{disscodepos}}

\newcommand{\Is}       {:=}

\newdimen\endofsize\endofsize=0.5em
\def\endofbeweis{~\quad\hglue\hsize minus\hsize
                 \hbox{\vrule height \endofsize width
\endofsize}\par}

\usepackage{amssymb}
\usepackage{algorithm}
\usepackage{algorithmic}
\usepackage{pdflscape}
\usepackage{numprint}
\npdecimalsign{.} 
\usepackage{xspace}
\usepackage{wrapfig}
\usepackage{graphics}
\usepackage{todonotes}
\usepackage{booktabs}
\usepackage{microtype}
\usepackage[left,pagewise] {lineno}
\definecolor {infocolor} {rgb} {0.6,0.6,0.6}

\def\MdR{\ensuremath{\mathbb{R}}}

\newcommand{\Id}[1]{\texttt{\detokenize{#1}}}

\newcommand{\ie}{i.\,e.,\xspace}
\newcommand{\eg}{e.\,g.,\xspace}
\newcommand{\etal}{et~al.\xspace}

\usepackage{color}

\newcommand{\strash}[1]{{\color{orange}[DS: #1]}}
\newcommand{\csch}[1]{{\color{blue}[CS: #1]}}
\newcommand{\lamm}[1]{{\color{blue}[SL: #1]}}
\newcommand{\sanders}[1]{{\color{blue}[PS: #1]}}
\newcommand{\werneck}[1]{{\color{blue}[RW: #1]}}
\renewcommand{\csch}[1]{}
\renewcommand{\strash}[1]{}
\renewcommand{\lamm}[1]{}
\renewcommand{\sanders}[1]{}
\renewcommand{\werneck}[1]{}

\def\comment#1{}

\def\withcomments{
  \newcounter{mycommentcounter}
   \def\comment##1{\refstepcounter{mycommentcounter}%
    \ifhmode%
     \unskip%
     {\dimen1=\baselineskip \divide\dimen1 by 2 %
       \raise\dimen1\llap{\tiny\bfseries \textcolor{red}{-\themycommentcounter-}}}\fi%
     \marginpar[{\renewcommand{\baselinestretch}{0.8}%
       \hspace*{3em}\begin{minipage}{5em}\footnotesize [\themycommentcounter]: \raggedright ##1\end{minipage}}]{\renewcommand{\baselinestretch}{0.8}%
       \begin{minipage}{5em}\footnotesize [\themycommentcounter]: \raggedright ##1\end{minipage}}}
  }

\definecolor{darkgreen}{RGB}{0,200,100}
\definecolor{orange}{RGB}{255,80,0}

\newcommand{\mytitle}{Better Process Mapping and Sparse Quadratic Assignment\footnote{This work was partially supported by DFG grants SA 933/11-1 and by the Austrian Science Foundation~(FWF, project P 31763-N31).}}

\begin{document}
\title{\mytitle}
\author[1]{Christian Schulz}
\author[2]{Jesper Larsson Träff}
\author[3]{Konrad von Kirchbach}
\affil[1]{University of Vienna, Faculty of Computer Science, Vienna, Austria}
\affil[2]{TU Wien, Faculty of Informatics, Vienna, Austria}
\affil[3]{University of Vienna, Faculty of Computer Science, Vienna, Austria}
\date{}

\Copyright{Christian Schulz, Jesper Larsson Träff, and Konrad von Kirchbach}
\maketitle

\begin{abstract}
Communication and topology aware process mapping is a powerful approach to reduce communication time in parallel applications with known communication patterns on large, distributed memory systems. We address the problem as a quadratic assignment problem (QAP), and present algorithms to construct initial mappings of processes to processors, and fast local search algorithms to further improve the mappings. By exploiting assumptions that typically hold for applications and modern supercomputer systems such as sparse communication patterns and hierarchically organized communication systems, we obtain significantly more powerful algorithms for these special QAPs.  Our multilevel construction algorithms employ perfectly balanced graph partitioning techniques and exploit the given communication system hierarchy in significant ways.  We present improvements to a local search algorithm of Brandfass \etal (2013), and further decrease the running time by reducing the time needed to perform swaps in the assignment as well as by carefully constraining local search neighborhoods. We also investigate different algorithms to create the communication graph that is mapped onto the processor network. Experiments indicate that our algorithms not only dramatically speed up local search, but due to the multilevel approach also find much better solutions~in~practice. 
\end{abstract}

\maketitle
\section{Introduction}
Communication performance between processes in high-performance parallel
systems depends on many factors. For example, communication is
typically faster if communicating processes are located on the same
processor node compared to the cases where processes reside on
different nodes.  This becomes even more pronounced for large
supercomputer systems where processors are hierarchically organized
into, \eg islands, racks, nodes, processors, cores with corresponding
communication links of similar quality.  Given the communication
pattern between processes and a hardware topology description that
reflects the strength of the communication links, one hence seeks to
find a good mapping of processes onto processors such that pairs of
processes exchanging large amounts of information are located closely.

Such a mapping can be computed by solving a corresponding quadratic
assignment problem (QAP) which is a hard optimization problem.  Sahni
and Gonzalez~\cite{SahniG76} have shown QAP to be strongly NP-hard
and, unless P=NP, admitting no constant factor approximation
algorithm.  In addition, there are no algorithms that can solve
meaningful instances of size $n$ with $n>20$ to optimality in a reasonable amount
of time~\cite{burkard1998quadratic}.  Hence, heuristic algorithms are
necessary in order to solve large scale instances.  Multiple
heuristics have been proposed to tackle real world
instances~\cite{brandfass2013rank,heider1972computationally,muller2013optimale}.
We present more details in Section~\ref{s:main}.

In this work, we make two important assumptions that are typically
valid for modern supercomputers and the applications that run on
those.  First, communication patterns are almost always sparse since
not all processes have to communicate with each other.  This is
especially true for large scale scientific simulations in which the
underlying models of computation and communication are already sparse, see,
\eg~\cite{catalyuerek1996dis,heuvelinecoop,schloegel2000gph}.
To efficiently parallelize the simulation one normally employs Graph
Partitioning (GP) techniques which then in turn yield a sparse
communication pattern between the processes.  Second, we assume that
the hardware communication topology under consideration is
hierarchical with communication links on the same level in the
hierarchy exhibiting the same communication speed. This is typically
the case for current high-performance systems.

Using these assumptions, we derive algorithms that are able to create high quality mappings, as well as faster local search algorithms for improving assignments. 
Overall, our algorithms are able to compute better solutions than other recent heuristics for the problem. Improving the (practical) complexity of such algorithms is highly important, since the number of cores available in supercomputers is still increasing dramatically. 
The rest of this paper is organized as follows.  
In Section~\ref{s:preliminaries}, we introduce basic concepts and describe relevant related work, such as the algorithm of Brandfass \etal~\cite{brandfass2013rank}, in more detail.
We present our main contributions in Section~\ref{s:main} and Section~\ref{s:modelcreation}. We also look at algorithms that create the communication model that has to be mapped. 
We implemented the techniques presented here in the graph partitioning framework KaHIP~\cite{kabapeE} (Karlsruhe High Quality Graph Partitioning). 
A summary of extensive experiments to evaluate algorithm performance is presented in Section~\ref{sec:experiments}, and indicate that our algorithm not only drastically speeds up local search, but due to the multilevel approach combined with high quality partitioning techniques also finds better solutions in practice. Lastly, using hierarchical multisection algorithms that take the system hierarchy into account for model creation further improves the results of the overall process mapping.

\vspace*{-.25cm}
\section{Preliminaries}
\vspace*{-.25cm}
\label{s:preliminaries}

The total communication requirement between the set of processes in (some
section of) an application can be modeled by a weighted communication graph. 
The underlying hardware topology can likewise be modeled by a weighted graph, but since the graph is complete (any physical
processor can communicate with any other physical processor through the underlying networks), we represent it by a topology cost matrix which 
can for instance reflect the costs of routing along shortest (cheapest) paths between processes.
Our abstract problem
is to embed the communication graph onto the topology graph under optimization criteria that we explain below. Throughout the paper, we assume 
that the number of nodes in host and topology graphs are the same.
Unless otherwise mentioned, a processing element (PE) typically represents a core of a machine.

In practice, often an input graph is given with a much larger number of vertices than the number of processors in the communication network. Assuming that edges of the input graph correspond to interaction between pairs of vertices, the problem is to assign the graph vertices to the processors of the communication network, such that the total communication induced by the vertices of the input graph is minimized, taking into account the hierarchical characteristics of the communication system. This can be viewed as a two stage process of first creating a smaller graph with as many vertices as there are processors in the communication network, and then mapping this smaller graph to the communication system. We refer to this approach as the \emph{model creation} problem, and the smaller graph as the \emph{model} for the given input graph.

\subsection*{Basic Concepts}
In the following, we consider undirected graphs $G=(V=\{0,\ldots, n-1\},E)$ with edge weights $\omega: E \to \MdR_{>0}$, node weights
$c: V \to \MdR_{\geq 0}$, $n = |V|$, and $m = |E|$.
We extend $c$ and $\omega$ to sets, \ie
$c(V')\Is \sum_{v\in V'}c(v)$ and $\omega(E')\Is \sum_{e\in E'}\omega(e)$.
We let
$\Gamma(v)\Is \setGilt{u}{\set{v,u}\in E}$ denote the neighbors of a node $v$.
A graph $S=(V', E')$ is said to be a \emph{subgraph} of $G=(V, E)$ if $V' \subseteq V$ and $E' \subseteq E \cap (V' \times V')$. We call $S$ an \emph{induced} subgraph when $E' = E \cap (V' \times V')$.  

Throughout the paper, $\mathcal{C}\in \MdR^{n \times n}$ denotes the communication matrix, and $\mathcal{D}\in \MdR^{n \times n}$ the topology or distance matrix. More precisely, $\mathcal{C}_{i,j}$ describes the amount of communication that has to be done between process $i$ and $j$, and $\mathcal{D}_{i,j}$ represents the weighted distance between PE $i$ and PE $j$.
That is, the cost for communicating the amount $\mathcal{C}_{i,j}$ between processors $i$ and $j$ is $\mathcal{C}_{i,j}\mathcal{D}_{i,j}$.
We follow Brandfass~\etal~\cite{brandfass2013rank} and others, and model the embedding problem as a quadratic assignment problem (QAP): Find a one-to-one mapping $\Pi$ of processes to PEs which minimizes the overall communication cost. More precisely, we want to minimize 
$J(\mathcal{C},\mathcal{D}, \Pi) := \sum_{i,j} \mathcal{C}_{\Pi(i), \Pi(j)}\mathcal{D}_{i,j}$ 
where the sum is over all PE pairs and $k=\Pi(i)$ means that process $k$ is assigned to PE $i$. Note that searching for the inverse permutation instead, \ie assigning process $i$ to PE $\Pi^{-1}(i)$, results in the same assignment problem as $\Pi$ is a one-to-one mapping. 
Throughout this work, we assume that $\mathcal{C}$ and $\mathcal{D}$ are symmetric -- otherwise one can create equivalent QAP problems with symmetric inputs \cite{brandfass2013rank}.
In this paper, we focus on \emph{sparse} communication patterns, and therefore do not want to store the complete communication matrix but instead represent it more efficiently as a graph. Furthermore, typical system topologies feature a hierarchy that can be exploit. For a given system, we assume that
hierarchy information, and in general $\mathcal{D}$, is given implicitly as part of the system description and can
be queried, and therefore does not have to be stored explicitly.

Graph partitioning is a key component in our algorithms to find initial solutions. 
The \emph{graph partitioning problem} looks for \emph{blocks} of nodes $V_1$,\ldots,$V_k$ 
that partition $V$, \ie $V_1\cup\cdots\cup V_k=V$ and $V_i\cap V_j=\emptyset$
for $i\neq j$. The \emph{balancing constraint} demands that 
$\forall i\in 1..k\gilt c(V_i)\leq L_{\max}\Is (1+\epsilon)\lceil c(V)/k \rceil$ for
some parameter $\epsilon$. 
In the \emph{perfectly balanced case} the imbalance parameter $\epsilon $ is set to zero, \ie no deviation from the average is allowed.
One commonly used objective is to minimize the total \emph{cut} $\sum_{i<j}\omega(E_{ij})$ where 
$E_{ij}\Is\setGilt{\set{u,v}\in E}{u\in V_i,v\in V_j}$. A vertex $v \in V_i$ that has a neighbor $w \in V_j, i\neq j$, is a boundary vertex. 
Another commonly used, similar objective is to minimize the maximum cut over all subsets. We do not consider this objective explicitly here.

\subsection*{Related Work}
There has been an enormous amount of research on GP, and we refer
the reader to \cite{GPOverviewBook,SPPGPOverviewPaper} for extensive
material and references.  All general-purpose methods that work well
on large real-world graphs are based on the multilevel principle.  The
basic idea can be traced back to multigrid solvers for systems of
linear equations \cite{Sou35}.  Well-known multi-level GP software packages
include Jostle~\cite{walshaw2000mpm}, Metis~\cite{karypis1998fast},
and Scotch~\cite{Scotch}. Jostle contains algorithms to compute
processor assignments in scientific simulations.  Jostle integrates
local search into a multi-level method to partition the model of
computation and communication. To do so, they solve the problem on the
coarsest level and afterwards perform refinement that takes the user
supplied network communication model into account. 
Scotch performs dual recursive bipartitioning to perform this task.

There is likewise a large literature on process mapping, often in the
context of scientific applications using MPI (Message-Passing
Interface).  Hatazaki~\cite{Hatazaki98} was among the first authors to
propose graph partitioning to solve the MPI process mapping problem
for unweighted process topologies. Tr\"aff~\cite{Traff02:topology} used
a similar approach, and gave one of the first non-trivial
implementations for the NEC vector supercomputers.  Mercier and
Clet-Ortega and later
Jeannot~\cite{mercier2009towards,MercierJeannot11} simplify the
mapping problem by only considering the topology inside the compute
nodes themselves and ignoring the topology of the network. Multiple
placement policies are investigated to enhance overall system
performance.  Yu et al.~\cite{YuChungMoreira06} discuss and implement
embedding heuristics for the BlueGene $3d$ torus systems.  Hoefler and
Snir~\cite{HoeflerSnir11} optimize instead the congestion of the
mapping, show that this problem is NP-complete, and give a corresponding
heuristic with an experimental evaluation based on application data
from the Florida Sparse Matrix Collection.  Routing aware mapping
heuristics taking the hierarchy of specific hardware topologies into
account were discussed in~\cite{Abdel-GawadThottethodiBhatele14}.
Vogelstein \etal~\cite{Vogelstein15} concentrate on solving general quadratic 
assignment and graph matching problems. They propose a gradient based heuristic
that involves solving assignment problems and give experimental evidence for
better solution quality and speed compared to certain other heuristics. The
worst-case complexity of their approach is high, $O(n^3)$ steps.

Previous work on model creation can be grouped into two categories. One line of research intertwines process mapping with graph partitioning. To this end, the objective of the partitioning algorithm -- most commonly the number of edges cut -- is typically replaced by an objective function that considers the processor distances. Throughout these algorithms, the distances have to be updated. 
The second category, which is the primary focus of our work, decouples partitioning and mapping. First, a graph partitioning algorithm is used to partition a large network into $n$ blocks, while minimizing some measure of communication, such as edge cut, and at the same time balancing the load (size of the blocks). Afterwards, a coarser model of computation and communication is created in which the number of nodes matches the number of PEs in the given processor network. This model is then mapped to a processor network of $n$ PEs with given pair-wise distances using a process mapping algorithm. 
We refer the reader to \cite{GPOverviewBook,SPPGPOverviewPaper} for more details.

\subsubsection*{Detailed Related Work}
\label{sec:relatedwork}
\label{s:related}
\label{s:detailedrw}
We now discuss related work by M\"uller-Merbach~\cite{muller2013optimale}, Heider~\cite{heider1972computationally} and Brandfass \etal~\cite{brandfass2013rank} as well as Glantz~\etal\cite{glantzMapping2015} 
in greater detail since our work either makes use of the tools proposed by those authors or because we compare against their results.  M\"uller-Merbach~\cite{muller2013optimale} proposes a greedy construction method to obtain an initial permutation for the QAP.
The method roughly works as follows: Initially compute the total communication volume for each processor and also the total distance for each core.
Note that this corresponds to the weighted degrees of the vertices in the communication and distance models, respectively. Afterwards, the process with the largest communication volume is assigned to the core with the smallest total distance. To build a complete assignment, the algorithm proceeds by looking at unassigned vertices and cores. For each of the unassigned processes the communication load to already assigned vertices is computed. For each core, the total distance to already assigned cores is computed. The process with the largest communication sum is assigned to the core with the smallest distance sum. Glantz~\etal\cite{glantzMapping2015} note that the algorithm does not link the choices for the vertices and cores and hence propose a modification of this algorithm called \emph{GreedyAllC} (the best algorithm in \cite{glantzMapping2015}). GreedyAllC links the mapping choices by scaling the distance with the amount of communication to be done. The algorithm has the same asymptotic complexity and memory requirements as the algorithm by M\"uller-Merbach. We also compare our proposed methods against GreedyAllC in Section~\ref{s:exp}.

Heider~\cite{heider1972computationally} proposes a method to improve an already given permutation/mapping. 
The method repeatedly tries to perform swaps in the assignment. To do so, the author defines a pair-exchange neighborhood $N(\Pi)$ that contains all permutations that can be reached by swapping two elements in $\Pi$. Here, swapping two elements means that $\Pi(i)$ will be assigned to processor $j$ and $\Pi(j)$ will be assigned to processor $i$ after the swap is done. The algorithm then looks at the neighborhood in a cyclic manner. More precisely, in each step the current pair $(i,j)$ is updated to $(i,j+1)$ if $j<n$, to $(i+1,i+2)$ if $j=n$ and $i < n-1$, and lastly to $(1,2)$ if $j=n$ and $i = n-1$. A swap is performed if it yields  positive gain, \ie the swap reduces the objective. The overall runtime of the algorithm is $O(n^3)$. We denote the~search~space~with~$N^2$.
To reduce the runtime, Brandfass \etal~\cite{brandfass2013rank} introduce a couple of modifications. First of all, only symmetric inputs are considered. If the input is not symmetric, the input is substituted by a symmetric one such that the output of the algorithm remains the same. Second, pairs $(i,j)$ for which the objective cannot change, are not considered. For example, if two processes reside on the same compute node, swapping them will not change the objective. Lastly, the authors partition the neighborhood search space into $s$ consecutive index blocks and only perform swaps inside those blocks. This reduces the number of possible pairs from $O(n^2)$ to $O(ns)$ overall pairs. We denote the search space with $\mathcal{N}_p$ (\emph{pruned neighborhood}).
In addition, instead of starting from the identity permutation, the authors use the method of M\"uller-Merbach~\cite{muller2013optimale} to compute an initial solution. This improves runtime of the local search approach as well as the objective~of~the~solution.


\section{Rank Reordering Algorithms}
\label{s:mainsection}
\label{s:main}

We now present our main contributions and techniques. This includes algorithms to compute initial solutions, speeding up the local search algorithms for sparse communication patterns and defining new search spaces for the local search algorithm. 
Throughout this section, we assume that the input communication matrix is already given as a graph $G_\mathcal{C}$, \ie no conversion of the matrix into a graph is necessary.
 More precisely, the graph representation is defined as $G_\mathcal{C}:=(\{1,\ldots, n\}, E[\mathcal{C}])$ where $E[\mathcal{C}] :=\{(u,v) \mid \mathcal{C}_{u,v} \not = 0\}$.
 In other words, $E[\mathcal{C}]$ is the edge set of the processes that need to communicate with each other. 
Note that the set contains forward and backward edges, and that the weights of the edges in the graph correspond to the entries in the matrix~$\mathcal{C}$. 

\subsection{Initial Solutions} 

\label{s:initialsolutions}
We propose two strategies exploiting the hierarchy. Intuitively, we
want to identify subgraphs in the communication graph of processes
that have to communicate much with each other and then place such
processes closely, \ie on the same node, same rack and so forth.  In
the following, we assume a homogeneous hierarchy of the supercomputer,
but our algorithms can be extended to heterogeneous hierarchies in a
straightforward way. Let $\mathcal{S}=a_1, a_2, ..., a_k$ be a
sequence describing the hierarchy of the supercomputer. The sequence
should be interpreted as each processor having $a_1$ cores, each node
$a_2$ processors, each rack $a_3$ nodes, \ldots,
such that the total number of processors is
  $n=\Pi_{i=1}^{k}a_i$.  We propose two algorithms to compute initial
mappings, a top down and a bottom up approach.  The first one,
\emph{top down}, splits the communication graph recursively and the
second one, builds a hierarchy \emph{bottom up}.

The \emph{top down approach} starts by computing a \emph{perfectly balanced} partition of $G_\mathcal{C}$ into $a_k$ blocks each having $n/a_k$ vertices (processes). The partitioning task is done using the techniques provided by Sanders and Schulz~\cite{kabapeE} which provide high quality partitions and guarantee that each block of the output partition has the specified amount of vertices. In principle, the nodes of each block will be assigned completely to one of the $a_k$ system entities. 
Each of the system entities provides precisely $n/a_k$ PEs. 
We then proceed recursively and partition each subgraph induced by a block into $a_{k-1}$ blocks and so forth. 
The recursion stops as soon as the subgraphs have only $a_1$ vertices left. 
In the base case, we assign processes to permutation~ranks. 

The \emph{bottom up approach} proceeds in the opposite order of the hierarchy. That means the communication graph $G_\mathcal{C}$ is split first into $k=n/a_1$ blocks with precisely $a_1$ vertices each. Again, this is done using the perfectly balanced partitioning techniques mentioned above. Each block will later on be assigned to a unique system entity that is able to host $a_1$ processes, \ie a node having $a_1$ cores. Then each of the blocks is contracted and we partition the contracted graph and so forth. In this case, if replacing
edges of the form $\set{u,w},\set{v,w}$ would generate two parallel edges $\set{x,w}$, we insert a single edge with $\mathcal{C'}_{x,w}=\mathcal{C}_{u,w}+\mathcal{C}_{v,w}$.  
This way, the correct sum of the distances are accounted for in later stages of the algorithm. 
The recursion stops as soon as the last hierarchy stage is reached, \ie the last graph with $n'$ vertices has been partitioned into $n'/a_k$ vertices with $a_k$ vertices each. 
Recall that vertices in the same block will be assigned to a specified subpart of the system. 
In this case, a vertex in the graph on the last level of the recursion represents a whole set of task with the property that the sum of the vertex weights of each block is precisely the amount of PEs that are present in the subsystem that they are assigned to. 
We then backtrack the recursion to construct the final mapping. 

\subsection{Faster Swapping}

Initially computing and later recomputing the objective function after a swap is performed is an expensive step in the algorithm of Brandfass \etal~\cite{brandfass2013rank}. 
In their work, both the communication pattern as well as the distances between the PEs are given as complete matrices. 
These matrices have a quadratic number of elements and hence the initial computation of the objective function costs $O(n^2)$ time.
After a swap is performed, Brandfass~\etal update the objective using the objective function value before the swap. 
This is done by looking at all elements in the corresponding columns of the communication and distance matrices. 
Overall, an update step in their algorithm takes $O(n)$ time which is clearly a bottleneck for sparse communication patterns.

We now describe how we speed up the initial computation as well as the update of the objective. 
As a first step, we rewrite the objective to work with the inverse of the permutation:
\begin{align*} 
J(\mathcal{C},\mathcal{D}, \Pi) &= \sum_{i,j} \mathcal{C}_{\Pi(i), \Pi(j)}\mathcal{D}_{i,j} \\
                                &=\sum_{u,v} \mathcal{C}_{u, v}\mathcal{D}_{\Pi^{-1}(u),\Pi^{-1}(v)}
\end{align*}
with the interpretation that task $u$ is assigned to PE $\Pi^{-1}(u)$. This makes it easier to work with the graph representation of the communication matrix.
We rewrite the objective to work with the graph representation instead of the complete communication pattern matrix~$\mathcal{C}$:
\[
J(\mathcal{C},\mathcal{D}, \Pi) := \sum_{(u,v)\in E[\mathcal{C}]} \mathcal{C}_{u, v}\mathcal{D}_{\Pi^{-1}(u),\Pi^{-1}(v)}.
\]
The first observation is that given an initial mapping, we can compute the initial objective in $O(n+m)$ time which is better for sparse graphs.  
Our next goal is to make the update of the objective fast after a swap has been performed.
To do so, let  
$\Gamma_{\Pi^{-1}}(u) := \sum_{v \in N(u)} \mathcal{C}_{u, v}\mathcal{D}_{\Pi^{-1}(u),\Pi^{-1}(v)}$
be the contribution to the objective of a single vertex $u$ given the current mapping. 
Note that by using $\Gamma_{\Pi^{-1}}$, we can again rewrite the objective 
$J(\mathcal{C},\mathcal{D}, \Pi) := \sum_{u \in V} \Gamma_{\Pi^{-1}}(u).$
Throughout the algorithm, the vertex contributions $\Gamma$ are always kept up to date. 
It is easy to see that performing a swap in the assignment only affects the nodes that are swapped themselves as well as their neighborhood in the communication graph. Hence, we only need to update the node contributions of those nodes and can update the objective accordingly.  
We update the node contributions as follows: Let $u$ and $v$ be the vertices to be swapped in their assignment $\Pi^{-1}$.
We start by subtracting the node contributions of all affected nodes from the objective. 
Before we perform the swap, we iterate over the neighbors of $u$ and $v$ and subtract the contribution induced by the edge connecting the neighbor from its $\Gamma$ value. 
We then set the node contributions of $u$ and $v$ to zero and perform the swap. Now we again iterate over all neighbors, basically recomputing the node contributions of $u$ and $v$, and at the same time adding the new contribution induced by the edge connecting the neighbor to its $\Gamma$ value.
As a last step, we add the new node contributions of all affected nodes from the objective. 
Overall, this takes $O(d_u+d_v)$ time where $d_u$ and $d_v$ are the degrees of the vertices $u$ and $v$ in the communication graph.

\subsection{Alternative Local Search Spaces}

We now define swapping neighborhoods using the communication graph $G_\mathcal{C}$. 
In the simplest version, assignments are only allowed to be swapped if the processes are connected by an edge in the communication graph, \ie the processes have to communicate with each other. 
We denote this neighborhood with $N_\mathcal{C}$. The size of the search space is $O(m)$ since it contains exactly $m$ pairs that may be swapped. 
Swaps are performed in random order. 
Local search terminates after $m$ unsuccessful swaps, \ie all pairs have been tried and no swap resulted in a gain in the objective.
Note that this approach assumes that swaps with positive gain are close in terms of graph theoretic distance in the communication graph. 
We also define augmented neighborhoods in which swaps are allowed if two processes have distance less than $d$ in the communication graph. We denote this neighborhood by $N^d_\mathcal{C}$. Note that this creates a sequence of neighborhoods increasing in size $N_\mathcal{C} \subseteq N^2_\mathcal{C} \subseteq \ldots \subseteq N^n_\mathcal{C} =  N^2$ where $N^2$ is the largest neighborhood used by Brandfass~\etal~\cite{brandfass2013rank} (see Section~\ref{s:detailedrw}).
Our experimental section shows that performing swaps with small graph theoretic distance in the communication graph is sufficient to obtain good solutions. 

\subsection{Miscellanea}
\label{s:miscell}

\subsubsection*{Constant Time Distance Oracle.} Storing the complete distance matrix requires $O(n^2)$ space. 
However, due to the problem structure it is not necessary to store the complete matrix. 
Instead one can build an interval tree over the PE given describing the hierarchy. 
The distance of two PEs can then be found by finding the lowest common ancestor in the tree.
Such a query can be answered in constant time by investing~$O(n)$~preprocessing~time~\cite{bender2000lca}.

We can use a simpler approach that obtains the distance of two PEs by a few, simple division operations. 
More precisely, for a hierarchy $\mathcal{S}=a_1, a_2, ..., a_k$ we initially build an array describing the sizes of the intervals on the different levels of the hierarchy. 
A query scans the implicitly given intervals from top to bottom until the PEs are not on the same subsystem, and then return the corresponding distance.  

\section{Model Creation} 
\label{s:modelcreation}

Recall the process mapping methodology: A graph partitioning algorithm is used to partition a large network into $n$ blocks, while minimizing some measure of communication, such as edge cut, and balancing the load. Afterwards, a coarser model of computation and communication is created. In this model each node corresponds to a block in the input network and edges are between nodes if there is an edge between the corresponding blocks of the input network. Edge weights in the graph model the amount of communication that needs to be done between the blocks. Note that the coarse graph corresponds to the communication graph $G_\mathcal{C}$ from the previous sections. This model is then mapped to a processor network of $n$ PEs with given pair-wise distances using a process mapping algorithm. 
The algorithm in this work map a model of computation and communication with $n$ vertices onto a processor network with $n$ PEs. Note that the identity mapping, \ie~the algorithm that maps task $i$ to process $i$ is also a possible option but the quality of this highly depends on the initial numbering of the blocks given by the graph partitioning algorithm. 
Also note that this process requires a graph partitioning algorithm. As we will investigate later, the way the partitioning algorithm operates has a large impact on the quality that can be obtained by using the identity mapping algorithm. In order to understand this fully, we now explain techniques used in a multilevel graph partitioning framework.

\subsection*{Multilevel Graph Partitioning}

Most, if not all, general-purpose methods that are able to obtain good partitions for large real-world graphs in reasonable time
are based on the multilevel principle~\cite{schloegel2000gph,GPOverviewBook,SPPGPOverviewPaper}. 
We now explain the multilevel graph partitioning approach implemented in KaHIP~\cite{kaffpa}.
Before we outline the multilevel approach, we need to define the notion of edge contractions. \emph{Contracting} an edge $\set{u,v}$ means to replace the nodes $u$ and $v$ by a new node $x$ connected
to the former neighbors of $u$ and $v$. 
We set $c(x)=c(u)+c(v)$ so that the weight of a node at each level is the number of nodes it is representing in the original graph. 
If replacing edges of the form $\set{u,w}$, $\set{v,w}$ would generate two parallel edges $\set{x,w}$, a single edge with
$\omega(\set{x,w})=\omega(\set{u,w})+\omega(\set{v,w})$ is inserted.
\emph{Uncontracting} an edge $e$ undoes its contraction. 

 The multilevel approach to graph partitioning consists of three main phases. 
In the \emph{contraction} (coarsening) phase, a hierarchy of graphs is created. 
There are multiple ways to do that. The most common way is to iteratively identify matchings $M\subseteq E$ and contract the edges in $M$. 
Contraction should quickly reduce the size of the input and each computed level should reflect the global structure of the input network. 
Contraction is stopped when the graph is sufficiently small to be directly partitioned using some expensive other algorithm.

In the \emph{local improvement} (or uncoarsening) phase, the matchings
are iteratively uncontracted.  Note that due to the way that the
contraction is defined, a partitioning of the coarse level creates a
partitioning of the finer graph having the same objective and balance.
After uncontracting a matching, a local improvement algorithm moves
nodes between blocks in order to improve the cut size or balance.
Usually variants of the Fiduccia-Mattheyses
algorithm~\cite{fiduccia1982lth} are used.  The intuition behind this
approach is that a good partition at one level will also be a good
partition on the next finer level, so that local search will quickly
find a good solution while moving only a very small amount of vertices
between the blocks.  Moving a node on a coarse level hierarchy usually
corresponds to the movement of a whole set of node movements of the
finest level of the hierarchy.  Intuitively, the multilevel scheme has
a global view on the optimization problem on the coarse levels of the
hierarchy and a very local view on the finest levels with respect to
the original graph.

\subsection*{Hierarchy Aware Model Creation}

There is an important detail: Systems like KaHIP and Metis~\cite{KarypisK98a} typically obtain a $p$-way partition by computing a $p$-way partition on the coarsest level through a recursive bisection strategy.
The graph is recursively divided into two blocks until the number of blocks is reached, \ie a bisection algorithm is used to split the graph into two blocks. 
More precisely, each bisection step itself uses a multilevel algorithm that stops as soon as the number of nodes is below an even smaller threshold for the number of nodes.
Greedy graph growing is used on the coarsest level to obtain a bipartition.
In KaHIP, if $k$ is not even, we split the graphs into two blocks, $V_1$ and $V_2$, such that $c(V_1) \leq (1+\epsilon) \lfloor \frac{p}{2}\rfloor  \lceil \frac{c(V)}{p} \rceil$ and $c(V_2) \leq (1+\epsilon)\lceil \frac{p}{2}\rceil \lceil \frac{c(V)}{p} \rceil$.
Block $V_1$ will be recursively  partitioned in $\lfloor \frac{p}{2} \rfloor$ blocks and block $V_2$ will be recursively partitioned in $\lceil \frac{p}{2} \rceil$ blocks.
We call this partitioning approach \emph{recursive bisection based model creation}.

It is important to note how the block IDs are distributed in this process. After the first bisection steps the first consecutive block IDs are assigned to the left hand side block and the remaining IDs are assigned to the right hand side block. Note that when computing the communication graph $G_\mathcal{C}$, node IDs correspond the block IDs in the original input. Hence, as observed in the experimental section, the identity mapping is good when elements in the system hierarchy parameter string $\mathcal{S}=a_1, a_2, ..., a_k$ are a power of 2. This is the case, because of the way the model creation/partitioning process works, the identity mapping yields good bisections in the model graph. On the other hand, if for example $a_k$ is not a power of two, then it is unlikely that the identity mapping corresponds to good partition of the model graph $G_\mathcal{C}$, as will be discussed again later.

With those observations in mind, we propose a different way to create the model graph. 
The approach is similar to the top down approach from Section~\ref{s:initialsolutions}. However, this time we use the approach to obtain an $p$-way partition of the input network.
Roughly speaking, instead of doing recursive bipartitioning of the input network, we now perform recursive multisection along the system hierarchy $\mathcal{S}$.
The approach starts by computing a partition of $G$ into $a_k$ blocks. 
We then proceed recursively and partition each subgraph induced by a block into $a_{k-1}$ blocks and so forth. 
The recursion stops after the last subgraphs in the recursion have been partitioned into  $a_1$ blocks. 
This algorithm also assigns consecutive block IDs recursively during the process to maintain locality.
Since local search algorithms typically move only few vertices on the higher levels in the multilevel hierarchy, the initial recursive structure is somewhat inherited to the final output partition.
Note that we now also have a partition from which we create a model $G_\mathcal{C}$, however, when we map the model onto our system hierarchy $\mathcal{S}$, the identity mapping intuitively is already quite good. 
We call this partitioning approach \emph{recursive hierarchical multisection based model creation}.


%
%

\section{Experiments}
\label{sec:experiments}
\label{s:exp}

\paragraph*{Methodology} 
We have implemented the algorithms described in Section~\ref{s:mainsection} within the KaHIP framework using C++ and compiled all algorithms using gcc 4.63 with full optimization's turned on (-O3 flag). 
We integrated our algorithms in the KaHIP v1.00 graph partitioning framework~\cite{kaffpa}.
The codes of Brandfass~\etal~\cite{brandfass2013rank} could not be made available to us, so that we implemented those algorithms in our framework as well. Our implementation also uses the sparse representation of the communication pattern.
GreedyAllC~\cite{glantzMapping2015} has been kindly provided by the authors.
We also compare against the dual recursive bisection codes of Hofler and Snir~\cite{HoeflerSnir11}~(LibTopoMap). 

\begin{wraptable}{r}{0.45\textwidth}
	\centering
        \small
 	\caption{Benchmark instance properties.}
 	\label{tab:test_instances_walshaw}
	\begin{tabular}{| l | r | r | }
			\hline
		 	Graph & $n$& $m$\\
		 	\hline \hline
		 	 \multicolumn{3}{|c|}{UF Graphs}\\
			\hline

		 	  cop20k\_A                                     & \numprint{99843}  & \numprint{1262244}\\
		 	  2cubes\_sphere                                & \numprint{101492} & \numprint{772886}\\
		 	  thermomech\_TC                                 & \numprint{102158} & \numprint{304700}\\
		 	  cfd2                                           & \numprint{123440} & \numprint{1482229}\\
		 	  boneS01                                        & \numprint{127224} & \numprint{3293964}\\
		 	  Dubcova3                                       & \numprint{146689} & \numprint{1744980}\\
		 	  bmwcra\_1                                      & \numprint{148770} & \numprint{5247616}\\
		 	  G2\_circuit                                    & \numprint{150102} & \numprint{288286} \\
		 	  shipsec5                                       & \numprint{179860} & \numprint{4966618}\\
		 	  cont-300                                       & \numprint{180895} & \numprint{448799}  \\
		 	
                          \hline
		 	  \multicolumn{3}{|c|}{ Large Walshaw Graphs}  \\
		 	
                          \hline
		 	  598a                                           & \numprint{110971} & \numprint{741934}   \\
		 	  fe\_ocean                                      & \numprint{143437} & \numprint{409593}   \\
		 	  144                                            & \numprint{144649} & \numprint{1074393}  \\
		 	  wave                                           & \numprint{156317} & \numprint{1059331} \\
		 	  m14b                                           & \numprint{214765} & \numprint{1679018}  \\
		 	  auto                                           & \numprint{448695} & \numprint{3314611}  \\
		 	
                          \hline
		 	   \multicolumn{3}{|c|}{ Large Other Graphs}\\
		 	
                          \hline
		 	  del23                                          & $\approx$8.4M     & $\approx$25.2M \\
		 	  del24                                          & $\approx$16.7M    & $\approx$50.3M \\
		 	
		 	  rgg23                                          & $\approx$8.4M     & $\approx$63.5M \\
		 	  rgg24                                          & $\approx$16.7M    & $\approx$132.6M\\
		 	
		 	  deu                                            & $\approx$4.4M     & $\approx$5.5M \\
		 	  eur                                            & $\approx$18.0M    & $\approx$22.2M \\
		 	
		 	  af\_shell9                                     & $\approx$504K     & $\approx$8.5M \\
		 	  thermal2                                       & $\approx$1.2M     & $\approx$3.7M   \\
		 	  nlr                                            & $\approx$4.2M     & $\approx$12.5M \\


		 	\hline
	\end{tabular}
        \vspace*{-1cm}
\end{wraptable}

Our experiments evaluate the objective of the quadratic assignment problem as well as the running time necessary to compute the solution.  
To keep the evaluation simple, we use mostly one system hierarchy configuration $\mathcal{S}, D$ which is specified in the corresponding chapter.
Depending on the focus of the experiment, we measure running time and/or the overall communication cost as defined in Section~\ref{s:preliminaries}.
We perform ten repetitions of each algorithm using different random seeds for initialization.
Unless otherwise mentioned, we use the geometric mean when reporting averages in order to give every instance the same influence on the \textit{final score}. 
The system we are using to compute solutions has four Octa-core Intel Xeon E5-4640 processors (32 cores) which run at a clock speed of 2.4 GHz. 
It has 512 GB local memory.

\paragraph*{Instances}

We use graphs from various sources to test our algorithm.
In Section~\ref{s:expqap}, we use these graphs as input to a partitioning algorithm that partitions them into a given number of blocks and then computes the communication graph $\mathcal{C}$ which is the input to our mapping algorithms.
For most of the experiments we use the recursive bisection based model creation approach. In Section~\ref{s:multisectionbasdedmodelcreation} we compare the results to the recursive multisection approach for creating the communication graph $G_\mathcal{C}$.
We use the largest six graphs from Chris Walshaw's benchmark archive~\cite{soper2004combined}.
Graphs derived from sparse matrices have been taken from the Florida Sparse Matrix Collection~\cite{UFsparsematrixcollection}. 
We also use graphs from the 10th DIMACS Implementation Challenge~\cite{benchmarksfornetworksanalysis} website. 
Here, \Id{rggX} is a \emph{random geometric graph} with
$2^{X}$ nodes where nodes represent random points in the unit square and edges
connect nodes whose Euclidean distance is below $0.55 \sqrt{ \ln n / n }$.
The graph \Id{delX} is a Delaunay triangulation of $2^{X}$ random points in the unit square. 
The graphs \Id{af_shell9}, \Id{thermal2},  and \Id{nlr} are from the matrix and the numeric section of the DIMACS benchmark set.
The graphs \Id{europe} and \Id{deu} are large road networks of Europe and Germany taken from~\cite{DSSW09}. 
Basic properties of the graphs under consideration can be found in Table~\ref{tab:test_instances_walshaw}.

\subsection{Sparse Quadratic Assignment Problem}

In this section, we look at the impact of the various algorithmic components that we presented throughout the paper. 
In general, we use a hierarchy $\mathcal{S}=a_1, \ldots, a_k$ describing the system hierarchy and communication parameters $D=d_1, \ldots, d_k$  describing the distances
between various cores in the subsystems. 
More precisely, $d_i$ describes the distance of 
two cores that are in the same subsystems for $i' < i$, and in different subsystems for $i' \geq i$. 
The total number of cores is then given by $n = \prod_i a_i$. 
Here, we focus on two different system configurations to keep the evaluation simple.
Our process in this section is as follows:
Take the input graph, partition it into $n$ blocks using the fast configuration of KaHIP, compute the communication graph induced by that (vertices represent blocks, edges are induced by connectivity between blocks, edge cut between two blocks is used as communication volume) and then compute the mapping of the communication graph to the specified system. 
\label{s:expqap}
\paragraph*{Speed-Up of Local Search} 

\begin{figure}[t]
\centering
\includegraphics[width=4.5cm,page=3]{plots/speedupplots.pdf}
\includegraphics[width=4.5cm,page=1]{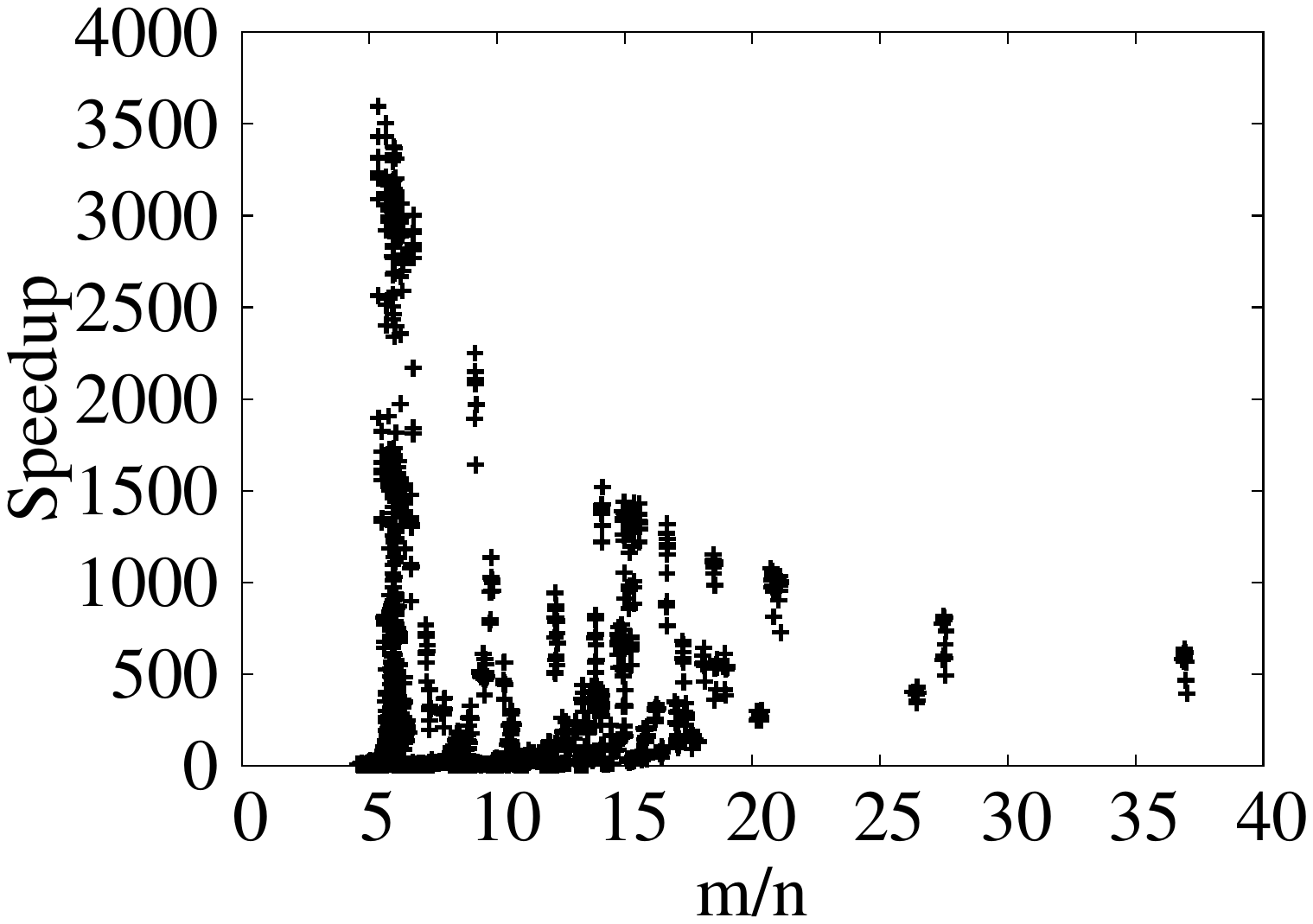}
\includegraphics[width=4.5cm,page=2]{plots/speedupplots.pdf}
\vspace*{-.15cm}
\caption{From left to right: Time of local search for both configurations (slow and fast), algorithmic speedup as a function of graph density, algorithmic speedup for the different instances.}
\vspace*{-.35cm}
\label{fig:speedup}
\end{figure}

We now take the algorithm configurations initially used by Brandfass \etal\cite{brandfass2013rank} and investigate the impact of our faster local search algorithms. 
The configurations are as follows: Use the greedy growing algorithm by M\"uller-Merbach (as described in Section~\ref{s:detailedrw}) to provide initial solutions and use the pruned local search neighborhood $N_p$ by Brandfass \etal~\cite{brandfass2013rank} (see Section~\ref{sec:relatedwork} for details). 
We run two configurations: One in which computing the gain takes linear time (the old algorithm) and one with our improved algorithm. 
In this experiment, we use $\mathcal{S}=4:16:k$, $D=1:10:100$ with $k = 2^i$, $i\in\{1, ..., 9\}$.
Note that the objective of the computed solutions by the algorithm using faster gain computations is precisely the same as their counter part, hence we do not report the value of the objective~in~this~section. 
The results of the experiments are summarized in Figure~\ref{fig:speedup} and Table~\ref{tab:runtimelssearch}. 
First, we observe that our new algorithm is \emph{always} faster than the old algorithm. 
This is expected, since 
the models of computation and communication that are mapped are indeed sparse. 
Table~\ref{tab:runtimelssearch} shows that our fast
local search algorithm scales almost linearly
\begin{wraptable}{r}{7cm}
\centering
\vspace*{-.25cm}
\small
\caption{Average running time and average speedup of local search for pruned search space $N_p$. $m/n$ is the average density of the generated instances, $t_\text{LS}$ the average running time of the algorithm using slow gain computations and $t_\text{fastLS}$ the average running time using fast gain computations.}
\label{tab:runtimelssearch}
\begin{tabular}{lr||r|r||r}
$n$ & $\overline{m/n}$ & $t_{\text{LS}}[s]$ & $t_\text{fastLS}$[s] & speedup\\
                \hline
 64  & 6.7  & 0.016               & 0.003   &5.3   \\
 128 & 7.3  & 0.064               & 0.006   &10.7  \\
 256 & 7.9  & 0.268               & 0.014   &19.1  \\
 512 & 8.3  & 1.073               & 0.029   &37.0  \\
 1K  & 8.8  & 4.263               & 0.059   &72.3  \\
 2K  & 9.2  & 17.083              & 0.124   &137.8 \\
 4K  & 9.7  & 68.360              & 0.260   &262.9 \\
 8K  & 10.3 & 268.907             & 0.540   &498.0 \\
 16K & 11.2 & \numprint{1075.107} & 1.158   &928.4  \\
 32K & 12.5 & \numprint{4348.374} & 2.472   &\numprint{1759.1} \\
\end{tabular}
\vspace*{-.25cm}
\end{wraptable}
in $n$ while the algorithm not using fast gain computations shows quadratic scaling behaviour.
The table also already shows a dependency of our algorithm on the density of the instances. 
This is due to the fact that the gain computation depends on the   
degrees of the vertices in the communication graph and is in alignment with our theoretical analysis.
The expected 
dependency on the density of the instances can also be seen more clearly in Figure~\ref{fig:speedup}.
The smallest algorithmic 
speedup obtained in this experiment is two and the largest speedup is approximately $\numprint{3596}$. 
We conclude that exploiting the sparsity of the application can improve the running time of local search significantly. We now always use fast gain computations.

\begin{figure}[t]
\centering
\includegraphics[width=5.5cm,page=1]{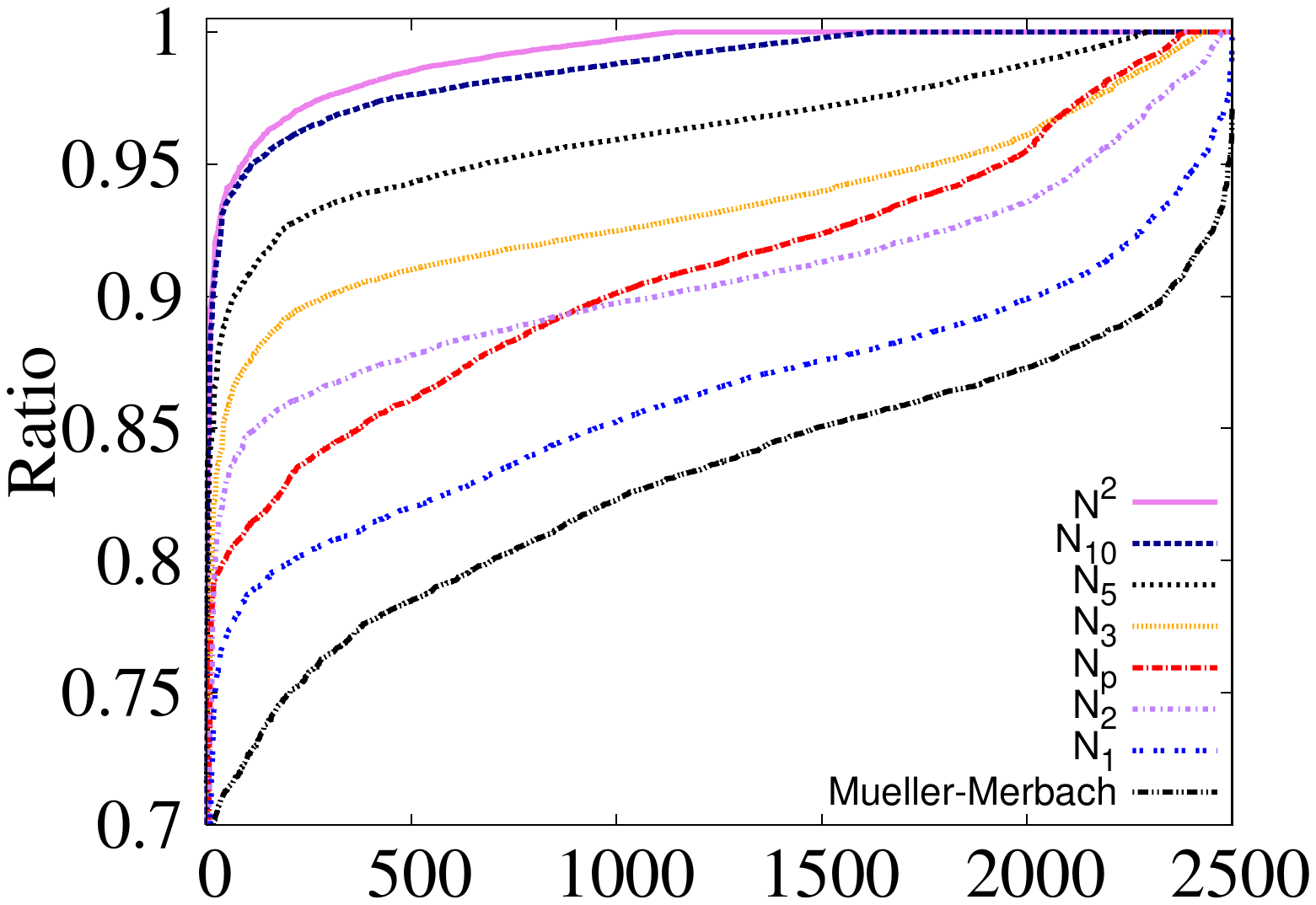}
\includegraphics[width=5.5cm,page=2]{plots/perfLSN.pdf}
\caption{Left/Right: performance plot w.r.t. solution quality/running time for different local search algorithms.}
\label{fig:perplots}
\end{figure}

\paragraph*{Local Search Neighborhoods} In this section, we look at the influence of local search neighborhoods on final solution quality. 
The base configuration used here employs the greedy growing algorithm by M\"uller-Merbach for initialization. Afterwards local search is done using the specified local search neighborhood, \ie the quadratic neighborhood $N^2$, the pruned quadratic neighborhood $N_p$ and the communication graph based neighborhoods $N_d := N_C^d$ for $d \in \{1,2,3,5,10\}$.  Again, we use $\mathcal{S}=4:16:k$, $D=1:10:100$ with $k = 2^i$, $i\in\{1, ..., 9\}$. 
To get a visual impression of the solution quality of the different algorithms, Figure~\ref{fig:perplots} presents performance plots using all instances. 
A curve in a performance plot for algorithm~X is obtained as follows: For each instance, we calculate the ratio between the objective or running time obtained by any of the considered algorithms and objective or running time of algorithm~X. These values are then sorted. Additionally, we present average ratios of solution quality and running~time~in~Table~\ref{fig:detailedlocalsearch}.
First, the local search algorithm using the $N_1$ neighborhood appears to be the fastest algorithm but also the worst in terms of solution quality. Compared to the initial construction heuristic it takes roughly a factor 1.34 in running time while improving solution quality by roughly 4\%.
With increasing distance $d$ for the local search neighborhood $N_d$, solutions improve but also the running time increases.  
As expected, the local search algorithm using the largest local search neighborhood $N^2$ computes the best solutions. Here, solutions generated by the initial heuristic are improved by roughly 20\%. However, this is also the slowest algorithm (a factor 443 slower than the initial construction heuristic).
Also note that we are only able to evaluate the performance of the algorithm at that scale due to the fast gain computations introduced in this paper.
Additionally, as $n$ increases the algorithm becomes much slower, as convergence of the algorithm takes more time for larger $n$. 
In contrast, the other local search neighborhoods show much better scaling behaviour as expected.  
The local search neighborhood $N_{10}$ is faster and computes solutions that are only slightly worse than $N^2$. 
For example, for $n=32$K the algorithm using $N_{10}$ is more than a factor nine faster and computes solutions that are only~5.5\%~worse.
\begin{table}[t]
\centering
\small
\caption{Average ratios for solution quality and running time. Baseline denotes the construction heuristic of M\"uller-Merbach without local search. Algorithms use the baseline algorithm and add local search with the respective local search neighborhood. 
        Comparisons are done against the baseline algorithm. Quality improvements are shown in \%.}
\label{fig:detailedlocalsearch}

\begin{tabular}{l||r|r|r|r|r||r|r|r|r|r}
$n$ & $N^2$ & $N_p$ &$N_1$&$N_2$& $N_{10}$ & $N^2$ & $N_p$ &$N_1$&$N_2$& $N_{10}$ \\
\hline
& \multicolumn{5}{l||}{baseline/\{baseline+local search\}} & \multicolumn{5}{l}{local search/baseline}\\
& \multicolumn{5}{l||}{quality improvement [\%]} & \multicolumn{5}{l}{average running time ratios: }\\
\hline
\numprint{64}  & \numprint{17.4} & \numprint{17.4} & \numprint{6.3} & \numprint{13.0}  & \numprint{17.2} &  \numprint{26.2}   & \numprint{27.1} & \numprint{2.6} & \numprint{13.3} & \numprint{44.1} \\
\numprint{128} & \numprint{16.0} & \numprint{10.9} & \numprint{3.8} & \numprint{8.5}   & \numprint{15.4} &  \numprint{63.9}   & \numprint{25.2} & \numprint{2.7} & \numprint{16.8} & \numprint{92.8} \\
\numprint{256} & \numprint{17.3} & \numprint{10.0} & \numprint{3.4} & \numprint{8.3}   & \numprint{17.3} &  \numprint{114.7}  & \numprint{18.9} & \numprint{2.5} & \numprint{16.3} & \numprint{149.0}\\
\numprint{512} & \numprint{17.6} & \numprint{8.9}  & \numprint{3.2} & \numprint{8.0}   & \numprint{17.5} &  \numprint{171.8}  & \numprint{11.3} & \numprint{1.8} & \numprint{12.7} & \numprint{190.2}\\
\numprint{1}K  & \numprint{18.8} & \numprint{8.2}  & \numprint{3.1} & \numprint{8.2}   & \numprint{18.2} &  \numprint{259.1}  & \numprint{6.8}  & \numprint{1.3} & \numprint{10.0} & \numprint{245.1}\\
\numprint{2}K  & \numprint{19.5} & \numprint{8.1}  & \numprint{3.1} & \numprint{8.2}   & \numprint{19.1} &  \numprint{348.2}  & \numprint{3.7}  & \numprint{0.9} & \numprint{7.0}  & \numprint{258.6}\\
\numprint{4}K  & \numprint{20.5} & \numprint{8.0}  & \numprint{3.3} & \numprint{8.7}   & \numprint{19.8} &  \numprint{472.0}  & \numprint{2.0}  & \numprint{0.6} & \numprint{5.1}  & \numprint{231.8}\\
\numprint{8}K  & \numprint{21.6} & \numprint{8.0}  & \numprint{3.6} & \numprint{9.4}   & \numprint{20.9} &  \numprint{728.2}  & \numprint{1.0}  & \numprint{0.5} & \numprint{4.0}  & \numprint{212.0}\\
\numprint{16}K & \numprint{23.1} & \numprint{8.3}  & \numprint{4.2} & \numprint{10.4}  & \numprint{22.1} &  \numprint{1030.8} & \numprint{0.6}  & \numprint{0.3} & \numprint{2.9}  & \numprint{173.6}\\
\numprint{32}K & \numprint{25.0} & \numprint{9.1}  & \numprint{5.4} & \numprint{11.9}  & \numprint{23.7} &  \numprint{1220.9} & \numprint{0.3}  & \numprint{0.2} & \numprint{2.1}  & \numprint{128.2} \\
        \hline
        \hline
overall: & \numprint{19.68} & \numprint{9.69} & \numprint{3.94} & \numprint{9.46} &  \numprint{19.12} & \numprint{443.58} & \numprint{9.69} & \numprint{1.34} & \numprint{9.02} & \numprint{172.54}  

\end{tabular}
\end{table}

\paragraph*{Initial Heuristics and Their Scaling Behaviour}
\begin{figure}[b]
\centering
\begin{minipage}{6.5cm}
\includegraphics[width=6.5cm]{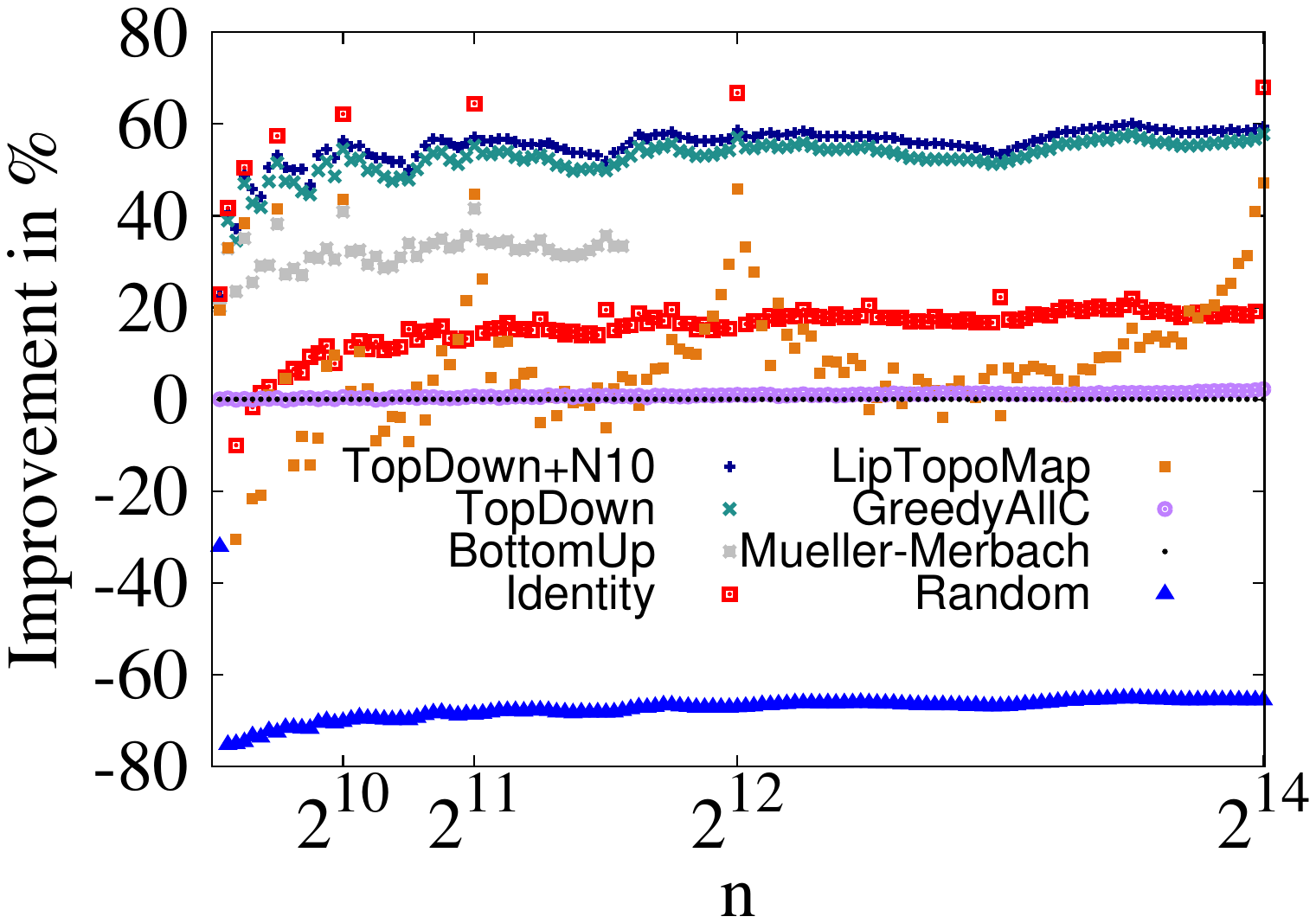} 
\end{minipage}
\begin{minipage}{6cm}
\vspace*{-.4cm}
\includegraphics[width=5.9cm]{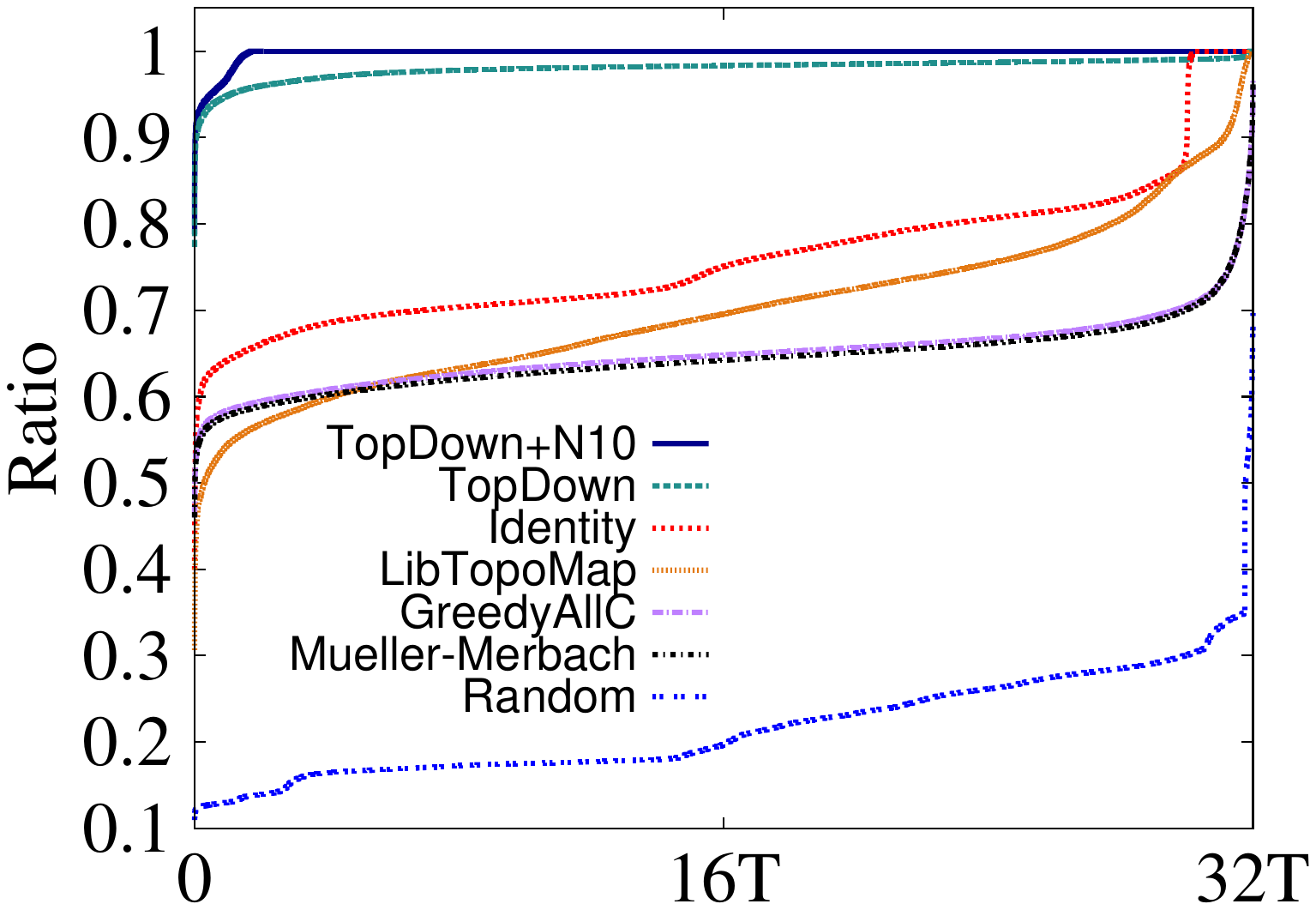}
\end{minipage}
\vspace*{-.4cm}
\caption{Average improvement in \% for different values of $n$ for different algorithms over the algorithm by M\"uller-Merbach (left) and a performance plot comparing solution quality (right). }
\label{fig:qualityofinitialheuristics}
\end{figure}
We now evaluate the different heuristics that can be used to create solutions. 
For the evaluation, we employ the algorithm of M\"uller-Merbach~\cite{muller2013optimale}, GreedyAllC~\cite{glantzMapping2015}, LibTopoMap~\cite{HoeflerSnir11} (dual recursive bisectioning), Identity, Random, the Bottom-Up as well as the Top-Down and the Top-Down algorithm combined with local search that uses the $N_{10}$ neighborhood (Top-Down+$N_{10}$). The problems we look at $\mathcal{S}=4:16:k$, $D=1:10:100$ with $k \in \{1, \ldots, 128\}$. We run the Bottom-Up algorithm only to $k\leq50$ due to its large running time. Figure~\ref{fig:qualityofinitialheuristics} shows the average improvement over solutions obtained by the algorithm of M\"uller-Merbach and a performance plot for the different algorithms.
Indeed, the random mapping algorithms performs worse than the algorithm of M\"uller-Merbach. On average, the objective computed by the algorithm is 67\% worse than the solutions computed by the algorithm of M\"uller-Merbach.
Our Top-Down algorithms yields the best solutions on most of the instances. On average, solutions computed by Top-Down are 52\% better than the solutions computed by M\"uller-Merbach. 
Adding local search with 
the $N_{10}$ neighborhood to the algorithm yields additional 5.3\% improvement on average.
GreedyAllC only improves slightly, \ie 1\% on average, over the algorithm of M\"uller-Merbach.
The identity mapping seems to be the best algorithm for powers of two. This is due to the way the input to the algorithms is constructed, \ie blocks are initially assigned by KaHIP. 
This algorithm uses a recursive bisection algorithm on the input graph to compute a model of computation and communication (the input to our mapping algorithms).
In each recursion it assigns consecutive blocks to the left side and to the right side. 
Hence, for powers of two, the identity mapping yields a strategy similar to using recursive bisection on the model to be mapped with good bisections.
If the number of elements is not a power of two, then the bisections implied by the identity are not good and hence it performs~worse. 

LibTopoMap is somewhere in between.  It mostly computes better solutions than the greedy algorithms but overall worse solutions than BottomUp and TopDown. On average, solutions are 8\% better than the solutions computed by the greedy algorithm of M\"uller-Merbach. Interestingly, its achieved solution quality is better when the number of vertices in the instances is close to a power of two. This is due to the fact that the algorithm uses dual recursive bisection on the communication and processor graph. However, when the input size is not close to a power of two, there are no good bisections in the processor graph.

In our experiments, Bottom-Up is the slowest algorithm. This is due to the fact that on the coarsest level large partitioning problems have to be solved.
The Top-Down algorithm does not have this problem, but is still slower than all other algorithms (except Bottom-Up). On average it is a factor 194 slower than the M\"uller-Merbach algorithm and a factor 40 slower than GreedyAllC. LibTopoMap is roughly a factor 18 slower than the algorithm of M\"uller-Merbach.
However, the running time of Top-Down is on average only 80\% of the time it takes to partition the input graph (using the fast configuration of KaHIP), \ie the time it takes to create the model which is the input to the mapping algorithms. Adding local search with the $N_{10}$ neighborhood to the algorithm costs additional time, on average 64\% of the time it takes to partition the graph.
Considering also the high solution quality advantage, we believe that the algorithms are still highly useful in practice.

\emph{Scalability.}
We now scale the problem size to $n=2^{19}$ processes/cores. We take the largest graph from our benchmark collection rgg24 and create mapping problems defined as $\mathcal{S}=4:16:128:k$, $D=1:10:100:1000$ with $k \in 2^i, i \in \{1, \ldots, 8\}$. We run Müller-Merbach and the TopDown+$N_{1}$ algorithm once. 
Both algorithms work well on our machine until $i=4$ $(n=2^{17})$, at which point there is not sufficient memory available if the implementations use the full distance matrix. Note that the machine has 512GB of memory.
Hence, we performed a second run of both algorithms computing distances online (as described in Section~\ref{s:miscell}).  
Note that the version of the Müller-Merbach algorithm is only able to solve larger problem sizes due to both of our changes: the sparse representation of the communication pattern as well as online computation of distances.
Computing distances online slows down Müller-Merbach roughly by a factor of five and local search by a factor of three.
The running time of TopDown remains the same since it uses the provided hierarchy instead of the distance matrix.
In turn the running time advantage of Müller-Merbach also decreases. This is also due to the fact that Müller-Merbachs algorithm is a quadratic time algorithm.
For the largest mapping problem ($n=2^{19})$, the Müller-Merbach algorithm takes a factor 1.64 longer than TopDown. 
Overall, computing distances online enables a potential user of the algorithms to tackle larger mapping problems.

\subsection{Multisection-based Model Creation}
\label{s:multisectionbasdedmodelcreation}
\begin{figure}[t]
\centering
        \vspace*{-.5cm}
        {\includegraphics[width=6.5cm,page=1]{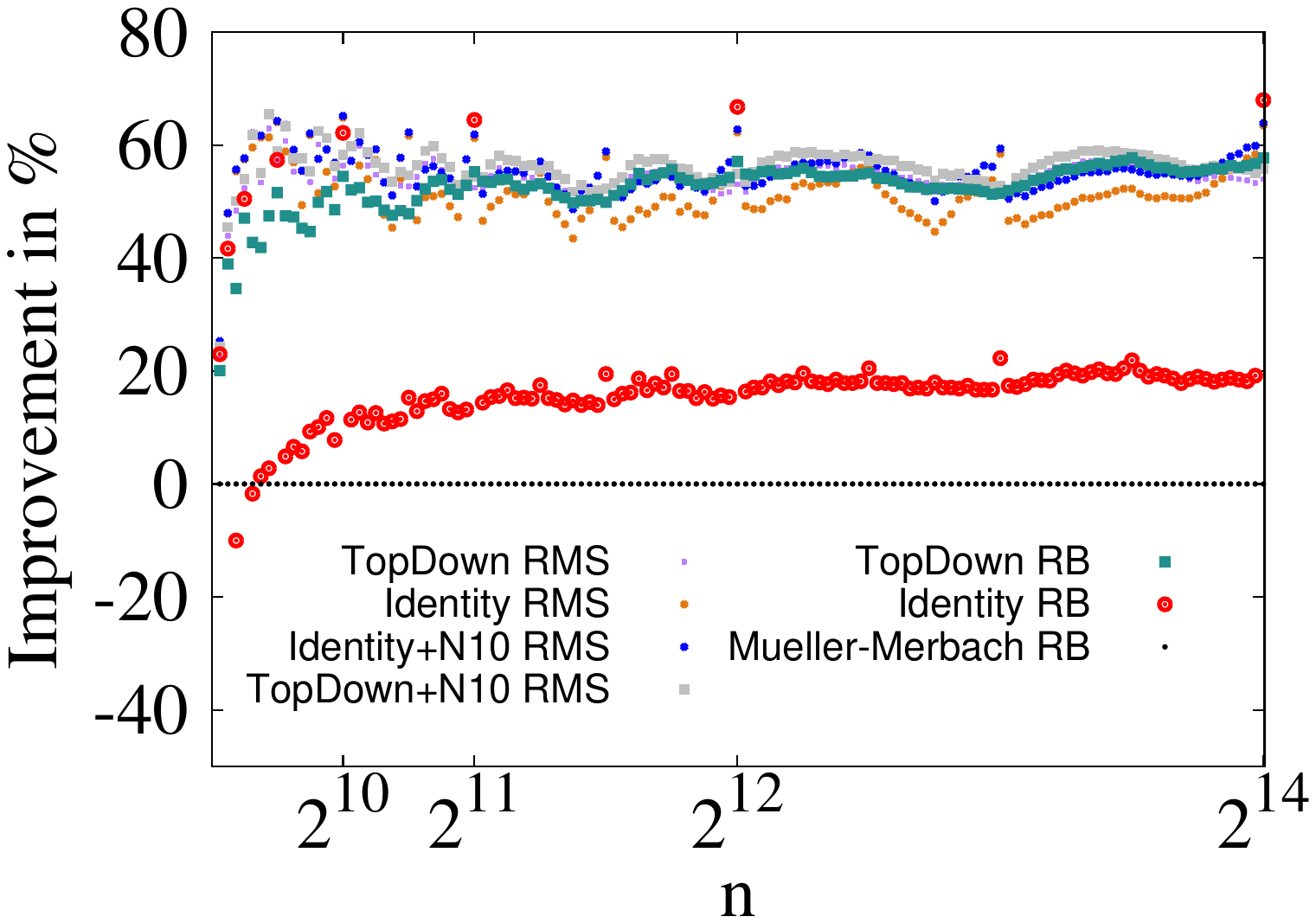}}%
        \vspace*{-.25cm}
        \caption{Average improvement in \% for different values of $n$ for different algorithms as well as different model creation algorithms. RB stands for a communication graph that has been obtained using the recursive bisection based algorithm and RMS obtained $G_\mathcal{C}$ by using the hierarchical multisection based approach that takes the system hierarchy into account.}
\label{fig:qap_vs_qap_kmodel}
\end{figure}

We now compare the different model creation algorithms. Recall that the model creation algorithm takes an input graph $G$ and partitions it using a graph partitioning algorithm.
From that partition we obtain the communication graph. In Section~\ref{s:modelcreation}, we presented two different strategies to perform the partitioning, the recursive bisection based algorithm (RB) as well as the hierarchical multisection based algorithm (RMS) that takes the system hierarchy into account.
The conjecture is that the employed strategy has an impact on the observed solution quality of the identity mapping and also on the overall mapping performance of the algorithms that map the communication graph $G_\mathcal{C}$ onto the processors.
Note that we now also compare 
the objective, $J(\mathcal{C},\mathcal{D}, \Pi) := \sum_{i,j} \mathcal{C}_{\Pi(i), \Pi(j)}\mathcal{D}_{i,j}$, for different $\mathcal{C}$ as different model creation algorithms create different communication models~$\mathcal{C}$ which are then mapped by our algorithms. However, from the application perspective this is unproblematic, since in some applications the user provides the input graph $G$ which needs to be both partitioned~\emph{and}~mapped.
We focus the evaluation on the best initial heuristic from the previous section, \ie we employ the Top-Down and the Top-Down algorithms combined with local search that uses the $N_{10}$ neighborhood (Top-Down+$N_{10}$) and additionally evaluate the Identity as well as the M\"uller-Merbach baseline algorithm. We use the abbreviations RB and RMS to indicate which algorithm has been used to create the communication graph~$G_\mathcal{C}$. The problems are defined as before: we look at $\mathcal{S}=4:16:k$, $D=1:10:100$ with $k \in \{1, \ldots, 128\}$.  Figure~\ref{fig:qap_vs_qap_kmodel} shows the average improvement over solutions obtained by the M\"uller-Merbach algorithm when RB is used to create $G_\mathcal{C}$. 

First of all, it can be seen that using the multisection strategy RMS that takes the system hierarchy into account drastically improves the quality of the identity mapping algorithm. 
While the Identity RB improves on M\"uller-Merbach RB by 18.2\%, it improves over M\"uller-Merbach RB by 51.6\% when the RMS model creation algorithm is used (Identity RMS). Moreover, it can be seen that the algorithm is now good for all values of $k$.
In contrast, TopDown RB improves over M\"uller-Merbach by 52.6\%. 
Hence, Identity RMS has comparable performance to TopDown RB. 
Identity RMS improves quality further to 55.4\% over M\"uller-Merbach RB when addition local search $N_{10}$ is used. 
However, also the quality of TopDown improves when the model creation algorithm is switched to RMS. 
In this case, TopDown RMS improves 54.1\% over M\"uller-Merbach RB and when additional local search $N_{10}$ is used, it yields the best algorithm with 56.1\% improvement.  

\section{Conclusion}

In high performance systems, different cores that are on the same processor usually have the same communication link quality when they communicate with each other, as do cores that are on the same node but not on the same processor~and~so~forth.
Using these assumptions, we derived algorithms to create initial mappings as well as faster local search algorithms with alternative local search spaces. Overall, our algorithms drastically speedup local search and are able to compute high quality solutions. 
Lastly, we have shown the impact of different model creation algorithms on the mapping algorithms. Using recursive multisection algorithms that take the system hierarchy into account improves the quality of the overall mappings achieved.

Important future work includes deriving distributed parallel algorithms for the problem.
Moreover, we want to investigate algorithms to create a hierarchy of the system if it is not provided as an input to our algorithm.
It may be worth to look at more complex local search neighborhoods, \eg local search spaces that allow to swap whole groups of assignments or allow swapping along cycles in the communication graph.
We also want to study the impact of our process mapping on parallel application performance.

{\bibliographystyle{plain}
\bibliography{phdthesiscs,additional}}
\end{document}